\documentclass[3p,times,twocolumn]{elsarticle}

\usepackage{ecrc}

\volume{00}

\firstpage{1}

\journalname{Nuclear Physics B Proceedings Supplement}

\runauth{G. Sottile et al.}

\jid{nuphbp}

\jnltitlelogo{Nuclear Physics B Proceedings Supplement}

\usepackage{amssymb}

\usepackage[figuresright]{rotating}

\begin{document}

\begin{frontmatter}



\dochead{}

\title{UVSiPM: a light detector instrument based on a SiPM
                sensor  working  in  single photon counting}

\author[label1]{G. Sottile}
\author[label1]{F. Russo}
\author[label1]{G. Agnetta}  
\author[label2]{M. Belluso}  
\author[label2]{S. Billotta}  
\author[label1]{B. Biondo}  
\author[label2]{G. Bonanno}  
\author[label1]{O. Catalano}  
\author[label1]{S. Giarrusso} 
\author[label2]{A. Grillo}  
\author[label1]{D. Impiombato} 
\author[label1]{G. La Rosa}  
\author[label1]{M.C. Maccarone} 
\author[label1]{A. Mangano}  
\author[label2]{D. Marano} 
\author[label1]{T. Mineo}  
\author[label1]{A. Segreto}  
\author[label1]{E. Strazzeri}  
\author[label2]{M.C. Timpanaro} 
\author{on behalf of the ASTRI Collaboration}

\address[label1]{Istituto di Astrofisica  Spaziale e Fisica Cosmica di Palermo, 
IASF-Pa/INAF, Palermo, Italy} 
\address[label2]{Osservatorio Astrofisico di Catania, OACT/INAF, Catania, Italy}

\begin{abstract}
UVSiPM is a light detector designed to measure the intensity of electromagnetic 
radiation in the 320--900 nm wavelength range. It has been developed in the framework 
of the ASTRI project whose main goal is the  design and construction
 of an end-to-end Small Size class Telescope prototype  for the Cherenkov 
Telescope Array. The UVSiPM instrument is composed by a multipixel Silicon Photo-Multiplier detector 
unit coupled to an electronic chain working in single photon counting mode with 10 
nanosecond double pulse resolution, and by a disk emulator interface card for 
computer connection. The detector unit of UVSiPM is of the same kind as the ones 
forming the camera at the focal plane of the ASTRI prototype. 
Eventually, the UVSiPM instrument can be equipped with a collimator to regulate its 
angular aperture.
UVSiPM, with its peculiar characteristics, will permit to perform several measurements 
both in lab and on field, allowing the absolute calibration of the ASTRI prototype.
\end{abstract}


\end{frontmatter}


\section{Introduction}
\label{}
UVSiPM is a stand-alone portable photon detector instrument designed to measure electromagnetic 
radiation in the 320--900 nm wavelength range. It has been developed in the framework of 
ASTRI (Astrofisica con Specchi a Tecnologia Replicante Italiana) Project \citep{ASTRI} a “Flagship Project”
 financed by the Italian Ministry of Education, University and Research (MIUR) and led by 
the Italian National Institute of Astrophysics (INAF). The ASTRI project is  focused,
in its first phase, on the design and construction  of an end-to-end Small Size scale Telescope prototype for 
the Cherenkov Telescope Array (CTA) \citep{CTA}, the international next generation ground-based 
observatory for very high energy gamma-rays 
(from a few tens of GeV up to several hundred of TeV).  The ASTRI project  will cover the 
highest part of the 
energy spectrum adopting innovative and challenging solutions never used in 
the framework of Cherenkov telescopes: the optics system will be based on a double mirror 
in Schwarzschild-Couder configuration instead of the standard single mirror; the camera 
 will be composed by an array of monolithic multipixel Silicon Photo-Multipliers 
(SiPM) instead of the usual Photo Multiplier Tubes (PMT). 

\begin{figure}
\centering
\includegraphics[width=7cm]{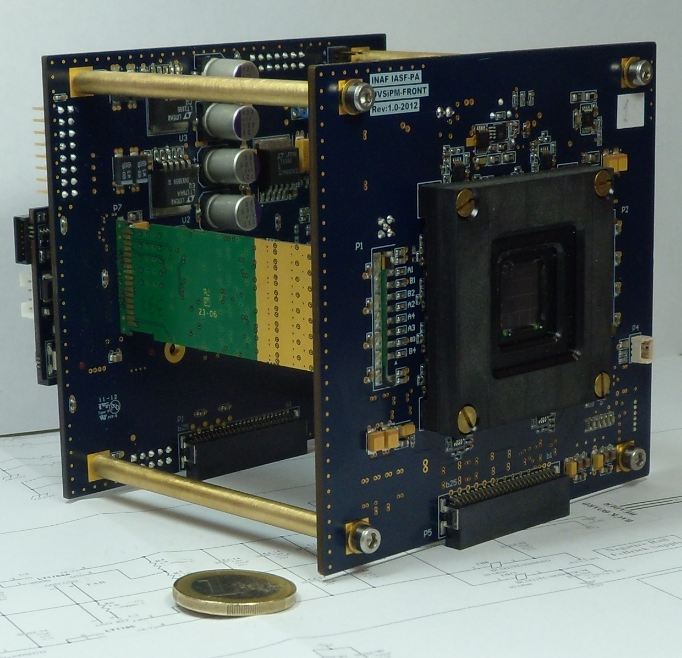}
\caption{UVSiPM detector.}
\label{fig1}
\end{figure}

UVSiPM,  developed at IASF-Palermo and  based on the same SiPM used for the ASTRI 
prototype, has the aim of characterizing the device and evaluating its performance in lab; the 
final goal is to use UVSiPM as tool for the on field absolute calibration of the  ASTRI prototype.
The instrument is designed around a single multipixel (4$\times$4 pixels) SiPM sensor coupled to
an electronic chain working in Single Photon Counting (SPC) mode, capable of 10 ns double pulse 
resolution.  Fig.~\ref{fig1} shows the UVSiPM 
detector and electronics. For on field measurements, a collimator will be mounted in front of 
the UVSiPM detector to regulate the angular aperture of the instrument.
 
\section{SiPM general characteristics}
\label{}
Silicon Photo-Multipliers are a family of light sensors with very interesting
characteristics. They are basically Avalanche Photo Diodes working in Geiger-mode, 
in which the reverse bias voltage is set beyond the Breakdown Voltage
(overvoltage). In this way, a single photon absorbed in Silicon develops a saturated
current avalanche with a gain of the order of 10$^6$.
There are many advantages in using SiPMs compared to the traditional 
PMTs: excellent single photon resolution, high Quantum
Efficiency (QE), no HV (bias voltages of the order of 30--90V), no damage when exposed
to ambient light, insensitive to magnetic fields, small size.
The drawbacks however are: high dark counts, afterpulses, optical crosstalk,
gain strongly dependent on temperature.

\begin{figure}
\centering
\includegraphics[width=7.5cm]{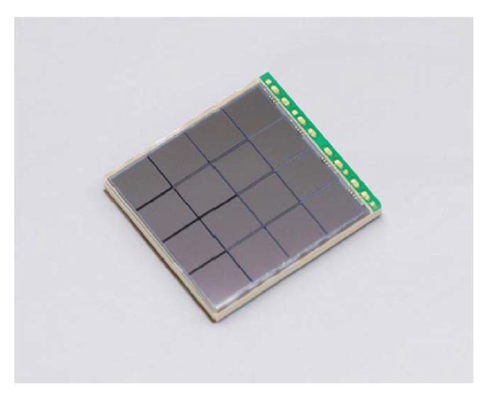}
\caption{The monolithic SiPM array Hamamatsu S11828-3344 used in the ASTRI camera.}
\label{fig2}
\end{figure}

SiPMs are the sensor that will be used to fill the camera at the focal plane of the ASTRI prototype.
The device chosen is the Hamamatsu S11828-3344M\footnote{http://sales.hamamatsu.com/info/eul/MPPC/MPPC.html}, 
a monolithic SiPM array in a configuration of 4x4 square pixels (see Fig.~\ref{fig2})
with a geometrical filling factor of about 74\%.
Each pixel, 3x3 mm in size, is made up of 3600 elementary diodes of 50 $\mu$m pitch. 

\section{The UVSiPM instrument}
\label{}
The UVSiPM instrument is basically composed by the SiPM sensor Hamamatsu
S11828-3344M, coupled to an electronic chain working in SPC
mode, with a double pulse resolution of 10 ns.

\begin{figure}
\centering
\includegraphics[width=7.5cm]{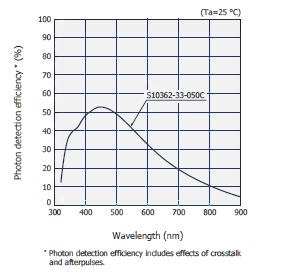}
\caption{A typical example of SiPM Photon Detection Efficiency (Courtesy Hamamatsu)}
\label{fig3}
\end{figure}

The total Photon Detection Efficiency (PDE) of SiPMs  is defined as:

$$PDE = QE \times FF \times GAP$$

\noindent
where $QE$ is the quantum efficiency, $FF$ the pixel filling factor and $GAP$ 
the Geiger avalanche probability. A typical PDE of an Hamamatsu SiPM 
is plotted in Fig.~\ref{fig3}. 
The sensor response ranges from UV to infrared wavelength due to the {\it p-on-n} 
technology used by Hamamatsu.  

Normally the PDE is measured from the mean value of the output current, when the device is exposed
to a known photons flux. In this way, pulses coming from crosstalk or afterpulses
are included in the total PDE leading to an overestimate of the real efficiency.
Evaluating the level of crosstalk and afterpulses is then one of
the task of the SiPM sensors characterization.

\begin{figure}
\centering
\includegraphics[width=7.5cm]{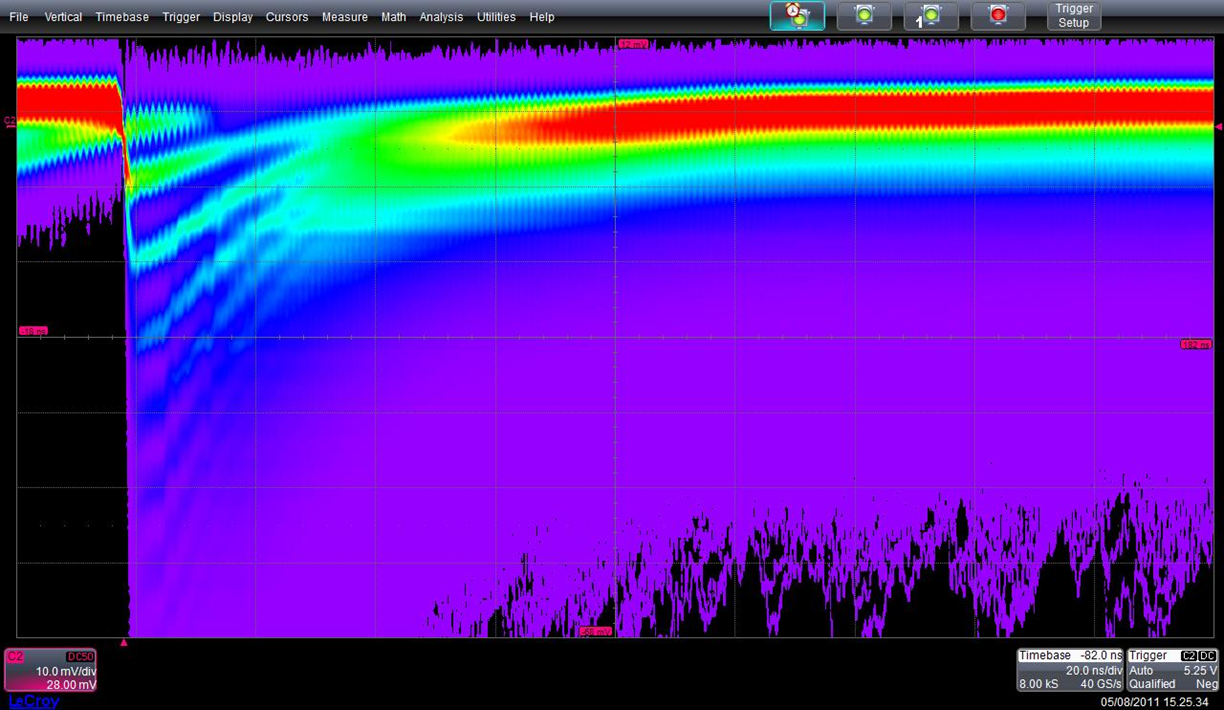}
\caption{A typical example of SiPM pulse waveform.}
\label{fig4}
\end{figure}

When a photon is absorbed in Silicon a current
avalanche produces an output pulse with a sharp rise-time (the discharge) 
followed by a very long tail (the recovery from the breakdown) as shown in 
Fig.~\ref{fig4}. 
Typical rise-time is of the order of few hundreds of picoseconds, whereas the
recovery time is in the range of tenths of nanoseconds.
From Fig.~\ref{fig4}, it is also evident the SiPM capability of distinguishing the
number of detected photoelectron from the level of the pulse height.
At high pulse rate pile-up effects occur and afterpulses also
contribute to rise the pile-up effects.

\begin{figure}[!h]
\centering
\includegraphics[width=7.5cm,height=4.0cm]{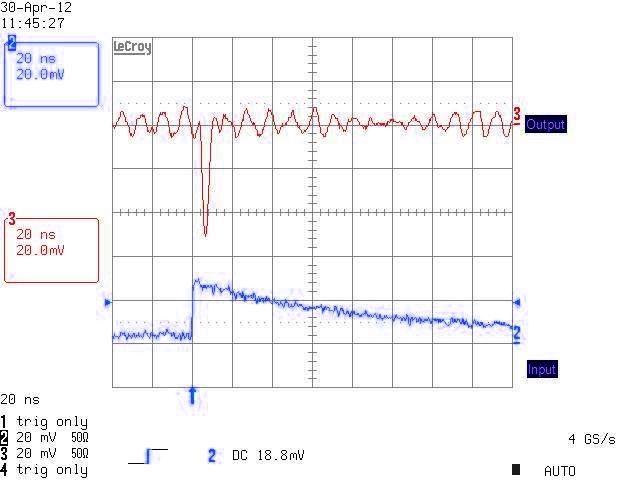}
\caption{Snapshot from the oscilloscope that shows an amplified 
SiPM-like signal (green curve) and the relative fast shaper output 
(brown curve).}
\label{fig5}
\end{figure}

\begin{figure}[!h]
\centering
\begin{minipage}[c]{7.5cm}
\includegraphics[width=7.0cm]{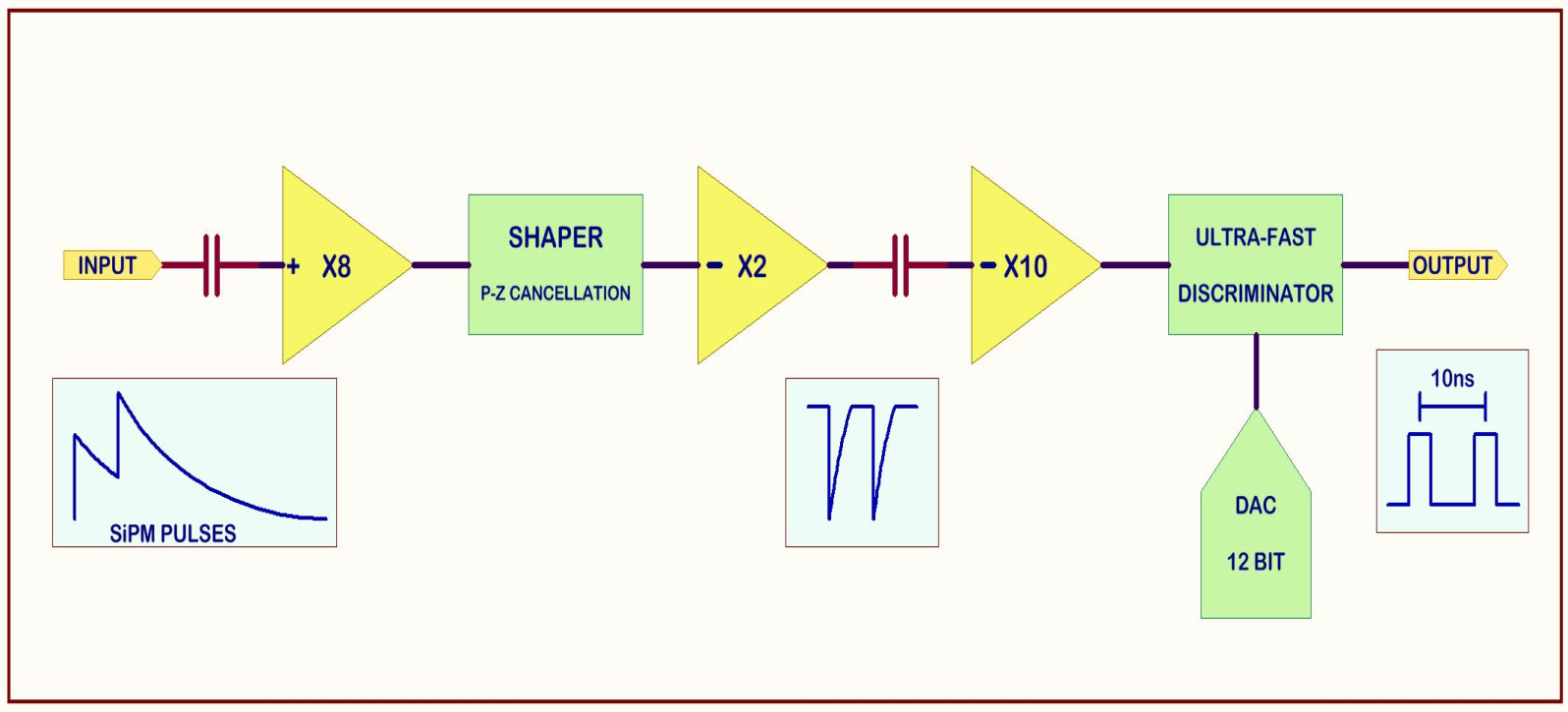}
\end{minipage}%
\hspace{2cm}%
\begin{minipage}[c]{7.5cm}
\includegraphics[width=7.0cm,height=4.0cm]{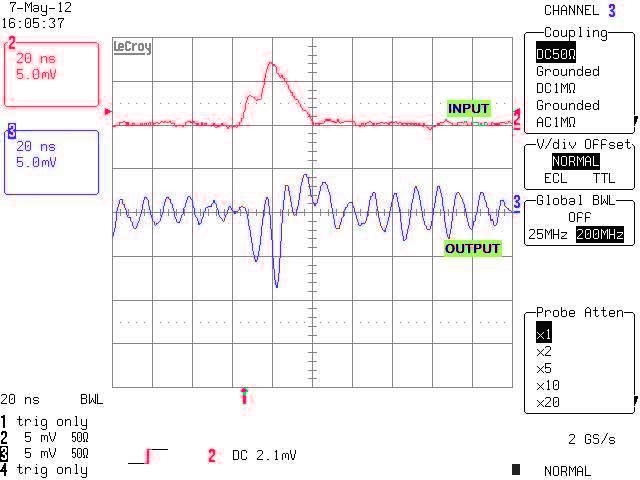}
\end{minipage}

\caption{The block diagram  of UVSiPM front-end electronics is presented in the top panel; 
a snapshot from the oscilloscope
in which two piled up pulses (green curve) are resolved (brown curve) is shown in the bottom panel.}
\label{fig6}
\end{figure}

The most efficient way to count pulses with the peculiar shape produced by the SiPM,
i.e. fast rise followed by a long tail, is to use  the derivative of the 
signal.
The electronics that performs this function is composed by an input amplifier buffer AC
coupled to the SiPM, followed by a shaping circuit that gives the derivative of the
signal cutting the negative part (pole-zero cancellation);  an inverting
amplifier drives a comparator whose threshold is set through a 12-bit DAC.
Fig. \ref{fig5} presents a snapshot from the oscilloscope that shows how a long tail pulse 
(SiPM-like) is changed into a short one.
Fig. \ref{fig6} shows the block diagram of the electronics and a snapshot from the oscilloscope
in which two piled up pulses taken from a pulse generator are resolved in two separate pulses.

A FPGA is in charge to manage the pulse counting and the housekeepings collection. The
communication with an external PC to control the instrument and the data
exchange is managed by the DELPHIN\footnote{http://www.iasf-palermo.inaf.it/cgi-bin/INAF/pub.cgi?href=facilities/electronic/instruments/delphin/index.html}, a proprietary interface board designed at IASF Palermo.

\section{UVSiPM and ASTRI: main applications}
 \label{}
UVSiPM has been designed as support instrumentation of the ASTRI prototype. 
Both of them use the same model of SiPM sensor and several kind of measurements, both 
in lab and on field, are foreseen for the characterization of the sensors and the calibration 
of the whole camera at the focal plane of the ASTRI prototype. 
The characterization of the SiPM will be carried out at the COLD Detector 
Laboratory in Catania (OACT/INAF). Among all the measurements, the most important are: 
PDE, dark counts, level of crosstalk and afterpulses  evaluated as function of
wavelength, overvoltage and temperature. 
Measurements of the detector response versus of the photon incidence angles are also
planned.

The design and implementation of UVSiPM derive from the well-tested UVscope instrument 
\citep{UVscope}, developed at IASF-Palermo, whose sensor is a calibrated Multi-Anode PMT
(MAPMT) working in SPC mode. The simultaneous usage of the two instruments on field, 
both provided with a calibrated John-B-25 Johnson/Bessel filter 
and equipped with collimators of proper length, will allow us to test the
performance of UVSiPM under real observing conditions.
The measurements will be based on the evaluation of the night sky background with the 
two instruments (UVSiPM and UVscope). The comparison will allow us to test the
end-to-end performances of SiPM vs MAPMT and then to use UVSiPM for the absolute 
calibration of  ASTRI telescope.


\bibliographystyle{elsarticle-num}



\end{document}